\documentclass[11pt]{article}

\usepackage[dvips]{epsfig}
\usepackage[dvips]{graphics}
\input{psfig.sty}

\def\be{\begin{equation}}
\def\ee{\end{equation}}
\def\bea{\begin{eqnarray}}
\def\eea{\end{eqnarray}}

\title{\bf Limits on the detectability of cosmic topology in  
                hyperbolic universes} 

\author{
G.I. Gomero$^{1}$\thanks{gomero@cbpf.br, chapita@gft.ucp.br}, \ 
M.J. Rebou\c{c}as$^1$\thanks{reboucas@cbpf.br}, \ 
R. Tavakol$^{1,2}$\thanks{r.tavakol@qmwul.ac.uk} \\ 
\\ 
$^{1}$  Centro Brasileiro de Pesquisas F\'\i sicas, \\
Rua Dr. Xavier Sigaud 150 \\
22290-180 Rio de Janeiro -- RJ, Brazil\\
\\
$^{2}$ Astronomy Unit, School of Mathematical Sciences, \\
Queen Mary, University of London, \\
Mile End Road, London E1 4NS, UK
}

\begin{document}

\date{} 

\maketitle

\begin{abstract} \noindent
We reexamine the possibility of the detection of the 
cosmic topology in nearly flat hyperbolic 
Friedmann-Lema\^{\i}tre-Robertson-Walker (FLRW) universes 
by using patterns repetition. We update and extend our 
recent results in two important ways: 
by employing recent observational constraints on the 
cosmological density parameters as well as the recent 
mathematical results concerning small hyperbolic 
$3$-manifolds. This produces new bounds with 
consequences for the detectability of the cosmic topology.
In addition to obtaining new bounds, we also give a concrete 
example of the sensitive dependence of detectability  
of cosmic topology on the uncertainties in the observational 
values of the density parameters.
\end{abstract}

\section{Introduction}   \label{intro}

Within the framework of standard cosmology, the universe 
seems to be well described by a locally homogeneous and 
isotropic Robertson-Walker (RW) metric
\begin{equation}
\label{RWmetric}
ds^2 = - c^2dt^2 + R^2(t)\,\{\, d\chi^2 + f^2(\chi)\,
       [\,d\theta^2 + \sin^2\theta\,d\phi^2\,] \,\} \;,
\end{equation}
where $t$ is a cosmic time, $f(\chi)= \chi\,,\;$ $\sin\chi\,,\;$ 
or $\sinh\chi\,,\;$ depending on the sign of the constant spatial 
curvature ($k = 0, \pm 1$), and $R(t)$ is the scale factor.
However, a RW metric does not uniquely specify the underlying RW
spacetime manifold $\mathcal{M}_4$, which can be decomposed 
into $\mathcal{M}_4 = \mathcal{R} \times M$. 
In traditional treatments of cosmology, 
the 3-space $M$ is usually taken to be one of 
the following simply-connected spaces: Euclidean  $E^3$, 
spherical $S^3$, or hyperbolic space $H^3$.
However, given that the simply-connectedness of our space $M$ 
has not been established  by cosmological observations, our 
3-space may equally well be any one of the possible quotient 
(multiply connected) manifolds $M = \widetilde{M}/\Gamma$, 
where $\Gamma$ is a discrete group of isometries of the 
covering space $\widetilde{M}$ acting freely on 
$\widetilde{M}$.
The action of $\Gamma$ tessellates the covering space 
$\widetilde{M}$ into identical cells or domains which 
are copies of what is known as fundamental polyhedron. 
An immediate observational consequence of a nontrivial topology 
(multiple-connectedness) of the 3-space $M$ is that the sky 
may show patterns repetition, i.e. multiple images of either 
cosmic objects or spots on the cosmic microwave 
background radiation, such as circles in the sky.

Questions of topological nature, 
such as whether we live in a finite or an infinite 
universe and what its shape may be are among the fundamental
open questions that modern cosmology needs to resolve. These 
questions go beyond the scope of general relativity (GR), since 
as a (local) metrical theory GR leaves the global topology 
of spacetime undetermined.

Given the wealth of increasingly
accurate cosmological observations, specially the 
recent observations of the cosmic microwave background
radiation (CMBR~\cite{BooMax,Jaffe}, these questions have become 
particularly topical (see, for example, 
\cite{CosmicTop}~--~\cite{ZelNov83}). 
It is therefore usually assumed that despite our present-day 
inability to \emph{predict\/} the topology  of the universe, 
it will become {\em detectable} as our observations
become more accurate.

An important outcome of the recent observations has been to 
suggest that the universe is almost flat (see, 
e.g., \cite{BooMax,Jaffe} and~\cite{Supernovae}~--~\cite{Acc}). 
This has motivated the recent study of the question of
detectability of the cosmic topology in such nearly-flat
FLRW universes~\cite{grt2001a}~--~\cite{EvLeLuUzWe}.
Here we update and extend our works~\cite{grt2001a,grt2001b}
by employing recent observational constraints on the cosmological 
density parameters as well as the recent mathematical results
concerning small hyperbolic $3$-manifolds.
In addition, we also find a concrete example of sensitive 
dependence of the detectable set of topologies on the 
observational bounds on the density parameters.

\section{Undetectability Indicators} \label{indic} 

Regardless of our present-day inability to predict the 
topology of the universe, its detection and determination 
is ultimately expected to be an observational problem. 
Recent studies have however shown that the near-flatness
of the universe, deduced from the recent analysis of 
observations data, may make the task of the detection of 
a possible nontrivial topology of the universe rather
difficult~\cite{grt2001a}~--~\cite{EvLeLuUzWe}. More precisely, 
it has been shown that if one uses patterns repetition, 
increasing number of nearly flat spherical and hyperbolic 
possible topologies for the universe become undetectable 
as $\Omega_0 \to 1$~\cite{grt2001a,grt2001b}.

The study of the possible non-trivial
topology of the spatial sections $M$
requires topological indicators which
could be put into correspondence with observations.
An intuitive starting point is
the comparison between the horizon radius and 
suitable characteristic sizes of the manifold $M$. 
A suitable characteristic size of $M$, which we shall use
in this paper, is the so-called injectivity radius $r_{inj}$ 
(radius of the smallest sphere `inscribable' in $M$), which 
is defined in terms of the length of the smallest closed 
geodesics $\ell_M\,$ 
by 
\begin{equation}  \label{rinj}
r_{inj} = \frac{\ell_M}{2} \,\;. 
\end{equation}
Using $r_{inj}$ we can define the indicator~\cite{grt2001a} 
\begin{equation}
\label{T_inj}
T_{inj}=\frac{r_{inj}}{\chi_{obs}} \; .
\end{equation}
Now, in any universe for which  $T_{inj} > 1$, every source
in the survey lies inside a fundamental polyhedron of $M$, no
matter what the location of the observer is within the manifold.
As a result there would be no repeated patterns in that survey 
and every method of the search for cosmic topology based on 
their existence will fail --- the topology of the universe is 
{\em undetectable\/} with this {\em specific\/} survey.
Now in practice, different surveys may be (and are often) employed.
There are three main surveys that can be used in the search for
repeated patterns in the universe: namely, clusters
of galaxies, containing clusters with redshifts of up to
$z_{cluster} \approx 0.3$; active galactic nuclei (mainly QSO's and
quasars), with a redshift cut-off of $z_{quasar} \approx 6$;
and maps of the CMBR with a redshift of $z_{cmb} \approx 1100$.
The crucial point is that the undetectability,
based on the employment of the above indicator~(\ref{T_inj}),
will be survey dependent. The latter survey, with
$z_{cmb} \approx 1100$
corresponding to the redshift of
the surface of last scattering, however, has
a unique place in practice, as it is in effect a limiting
survey with the deepest depth.
Thus the
quotient~(\ref{T_inj}) computed with $z_{cmb}$
gives the lowest observational bound in practice for the 
indicator $T_{inj}$.
At a theoretical level, on the other hand, an absolute 
lower bound is given by the indicator defined in terms 
of the horizon radius,
\begin{equation} 
\label{T_hor}
T_{inj}^{hor}=\frac{r_{inj}}{\chi_{hor}} \;.
\end{equation}

The undetectability which arises from the
condition $T_{inj}^{hor} \geq 1$ is obviously
{\em survey independent\/}, and when this
inequality holds no multiple images (or patterns repetition) 
will arise from any survey, including, of course, CMBR. 
Thus, any method for the search of cosmic topology based 
on the existence of repeated patterns will fail --- the 
topology of the  universe is {\em definitely undetectable\/} 
in such cases.

It is worth emphasizing  that the indicator $T_{inj}$ is 
useful for the identification of cosmological models whose 
topology is undetectable through methods based on 
the presumed existence  of multiple images, for when 
$T_{inj} \geq 1$, the whole region covered by a specific 
survey lies inside a fundamental polyhedron of $M$. However, 
without further considerations, nothing can be said about 
the detectability when $T_{inj} < 1$. In fact, in this case, 
even if the radius of the depth of a given survey is larger 
than $r_{inj}$, it may be that, due to the location of the 
observer, the whole region covered by the specific survey 
would still be inside a fundamental polyhedron of $M$, 
making the topology undetectable. 
This is the case when the smallest closed geodesic 
that passes through the observer is larger than 
$2 \chi_{obs}$.%

In section~\ref{detect} we shall use $T_{inj}^{hor}$ and 
the indicator $T_{inj}$ to examine the detectability of 
set of small hyperbolic universes in the light of the most 
recent observations.

\section{Hyperbolic 3-manifolds} \label{hyperman}

In this section we shall briefly recall some relevant 
facts about hyperbolic $3$-manifolds which will be 
usefull in the following section. 
We note in passing that in line with the usual mathematical 
practice in investigations of hyperbolic manifolds, we shall
use the curvature radius as the unit of length. 

Despite the enormous advances made in the last few decades,
there is at present no complete classification of hyperbolic
$3$-manifolds. However, a number of important results have been
obtained. Here we shall briefly recall a number of  results 
concerning closed orientable hyperbolic $3$-manifolds
which will be useful for our purposes in this work:
\begin{enumerate}
\item
Mostow's rigidity theorem~\cite{Mostow}, which ensures a 
rigid connection between
geometrical quantities and topological features
in hyperbolic $3$-manifolds. 
Thus once the topology is specified, all metrical quantities,
such as the volume and the lengths of their closed geodesics 
are topological invariants for a given $3$-manifold.
We note, however, that the volume alone does not
uniquely specify the $3$-manifold, and consequently there are 
topologically distinct hyperbolic $3$-manifolds with
the same volume.
\item
Compact orientable hyperbolic $3$-manifolds constitute 
a countable infinity of countably infinite 
number of sequences, ordered according to their volumes. 
Moreover, a fixed sequence has an accumulation of compact 
manifolds near a limiting volume set by a cusped 
manifold, which has finite volume, is non-compact, 
and has infinitely long cusped corners~\cite{Thurston82}.
\item
According to a result of Thurston~\cite{Thurston82},
there exists a hyperbolic 3-manifold with a minimum 
volume. This has very recently been shown by  
Agol~\cite{Agol} to be greater than $0.32095$, 
improving an earlier bound ($0.2815$1) by 
Przeworski~\cite{Przeworski};
\item
Closed orientable hyperbolic $3$-manifolds can be 
constructed and studied with the publicly available 
software package SnapPea~\cite{SnapPea} (see 
also~\cite{AdamsSnap}). The compact manifolds are
constructed through a so-called Dehn surgery which
is a formal procedure identified by two coprime 
integers, i.e. winding numbers $(n_1,n_2)$.  
SnapPea names manifolds according to the seed cusped
manifold and the winding numbers. So, for example, the
smallest hyperbolic manifolds is named as
$m003(-3,1)$, where $m003$ corresponds to a 
seed cusped manifold, and $(-3,1)$ is a pair of
winding numbers.
\item
There is a census by Hodgson and Weeks~\cite{SnapPea,HodgsonWeeks} 
containing $11031$ orientable closed hyperbolic 3-manifolds 
ordered by increasing volumes. Besides the volumes, it also 
provides other information, such as the
solution type, the length of shortest closed 
geodesic and the first fundamental group. 
The smallest (volume) manifold in this census 
(Weeks' manifold) has volume $\mbox{Vol}(M) = 0.94271$, 
and is conjectured to be the hyperbolic 3-manifold with
minimum volume. But, as was mentioned above,
the best current estimate for the volume  $V$ of the 
smallest closed hyperbolic orientable $3$-manifold is 
that in lies in the range $ 0.32095 < V \le 0.94271$.
\item
Clearly, there is a lower bound on the lengths of 
geodesics in any finite set of small volume closed 
orientable hyperbolic $3$-manifolds. More importantly,
according to a very recent theorem of Hodgson and 
Kerckhoff~\cite{HodKer} the shortest geodesic in 
closed orientable hyperbolic 3-manifolds with volume 
less than 1.7011 must have length greater than $0.162$, 
corresponding to a lower bound on $r_{inj}$ of $0.081$. 
We recall that there are 19 manifolds in Hodgson and Weeks
census with volume smaller than $1.7011$. We also 
note that the closed census intentionally excludes
all manifolds containing geodesics of length less than 
$0.300$, which means that the lower bound of $0.081$
may in fact correspond to a larger set of manifolds. 
It is worth noting that this is an important 
improvement on the lower bound of $0.09$
due to Przeworski~\cite{Przeworski} which we used 
in our previous work~\cite{grt2001b}. 
That bound was established for a set of  manifolds 
with volumes less than 0.94274, which only contained 
one known manifold, namely the Weeks' manifold. 
\end{enumerate}

\section{Detectability and observations} \label{detect}
A combination of recent independent astrophysical and 
cosmological observations seems to indicate that we live 
in an accelerating FLRW universe with nearly f\/lat spatial
sections (with $\Omega_0 \simeq 1$), which contains about 
$\sim 30 \%$ dark matter, close to $\sim 70 \%$
dark energy together with a small amount of 
baryonic matter of the order of few percent (see, 
for example, references~\cite{BooMax,Jaffe} and 
\cite{Supernovae}~--~\cite{Acc}).

In the light of these observations, we assume that the
universe can be locally described by a FLRW 
metric~(\ref{RWmetric}), and that the matter content of 
the universe is well approximated by dust of density $\rho_m$ 
plus a cosmological constant $\Lambda$. The Friedmann 
equation is then given by
\begin{equation} 
\label{Feq}
H^2 =\frac{8 \pi G \rho_m }{3} -\frac{k c^2}{R^2} 
            +\frac{\Lambda\,c^2}{3}\;,
\end{equation}
where $H=\dot{R}/R$ is the Hubble parameter and $G$ is 
Newton's constant. Introducing 
$\Omega_m = \frac{8 \pi G \rho_m}{3 H^2}\,$ 
and 
$\,\Omega_{\Lambda} \equiv \frac{8\pi G \rho_{\Lambda}}{3 H^2}= 
\frac{\Lambda\,c^2}{3 H^2}\,$,  
and letting
$\,\Omega = \Omega_m + \Omega_{\Lambda}$, 
equation~(\ref{Feq}) gives
\begin{equation}   
\label{R-eq}
H^2 R^2 (\Omega - 1) = k c^2 \;.
\end{equation}
{}From Eq.~(\ref{R-eq}), for hyperbolic models 
($\Omega_0 < 1$), the redshift-distance relation in 
units of the curvature radius, $R_0$, reduces 
to~\cite{grt2001a} 
\begin{equation} \label{redshift-distR}
\chi\,(z) = \sqrt{|1-\Omega_0|} 
   \int_0^{z} \left[ (1+x)^3 \Omega_{m0} + 
 \Omega_{\Lambda 0} -(1+x)^2 (\Omega_0 -1)\right]^{-1/2} dx\;,
\end{equation}
where the subscript $0$ denotes evaluation at present time.
The horizon radius $\chi_{hor}$ is defined 
by~(\ref{redshift-distR}) for $z=\infty$.
Written in this form the redshift-distance relation is very 
convenient for the study of hyperbolic universes, since 
the curvature radius is used as the unit of length.

\begin{figure}[!htb]
\centerline{\def\epsfsize#1#2{0.5#1}\epsffile{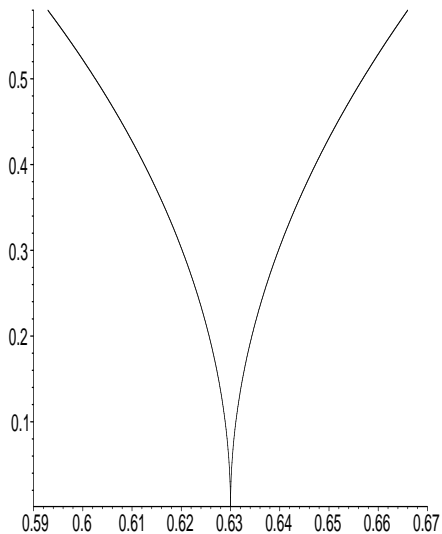}}
\caption{\label{fig=0.01} The behaviour of the horizon radius $\chi_{hor}$
in units of curvature radius, for FLRW models with $\rho_m$ and 
$\Lambda$ as a function of the density parameters 
$\Omega_{\Lambda}$ for $\Omega_m\,= 0.37$,
which is the middle value for $\Omega_m\,$. 
This figure shows clearly the rapid way $\chi_{hor}$ falls 
off to zero for nearly flat hyperbolic universes,
as $\Omega_0 = \,\Omega_{m0} + \Omega_{\Lambda 0} \,\,\to 1$.
The vertical   represents $\chi_{hor}$, while the horizontal 
axis gives $\Omega_{\Lambda}\,$.} 
\end{figure}

To begin with, we recall that the chances of 
detecting the topology of a nearly flat compact universe 
from cosmological observations become smaller as 
$\chi_{hor} \to 0\,$  ($\chi_{hor} \ll R_0$). 
Thus as a first step
in studying the constraints on detectability we 
consider the horizon radius function 
$\chi_{hor}(\Omega_{m0},\Omega_{\Lambda0})\,$ given 
by~(\ref{redshift-distR}) with $z=\infty$, for a typical 
fixed value $\Omega_{m0}= 0.37$, which 
is the middle value of the bounds on $\Omega_{m0}\,$,
obtained recently~\cite{Jaffe} by combining measurement 
of the CMBR anisotropy (BOOMERANG-98, MAXIMA-1 and COBE 
DMR) together with supernovae Ia (SNIa) and large scale 
structure (LSS) observations. 
Figure~\ref{fig=0.01} shows the behaviour of $\chi_{hor}$ as 
a function of $\Omega_{\Lambda 0}$ for this fixed value of 
$\Omega_{m0}\,$. 
Clearly, the limiting case of flat universes
($\Omega_0=1$) corresponds to the point at which the curves 
touch the horizontal axis. 
This figure clearly demonstrates the rapid way $\chi_{hor}$ 
drops to zero in a narrow neighbourhood of the   
$\Omega_0 = 1$. From the observational point of view,
this shows that the detection of the topology of the 
nearly flat hyperbolic universes becomes more 
and more difficult as $\Omega_0 \to 1$, a limiting 
value favoured by recent observations.

To obtain more quantitative information, we employ
the indicator $T_{inj}$ to examine the detectability of cosmic
topology of hyperbolic universes with nontrivial topologies.
Given the existence of more recent estimates of 
the cosmological density parameters, we shall update and extend 
our previous results by considering, in addition to  
the hyperbolic sub-interval given by
Bond {\em et al.\/}~\cite{Bond-et-al-00a}: 
\begin{equation}
\label{hyper-data1}
\Omega_0 \in [0.99,1) \qquad \mbox{and} \qquad
\Omega_{\Lambda 0} \in [0.63,0.73]
%
\end{equation}
the hyperbolic sub-interval consistent with a more recently 
bound on the density parameters given by Jaffe 
{\em et al.\/}~\cite{Jaffe}
\begin{equation}
\label{hyper-data2}
\Omega_0 \in [0.98,1) \qquad \mbox{and} \qquad
\Omega_{\Lambda 0} \in [0.62,0.79] \;.
%
\end{equation}

To make a comparative study, we consider each set of these 
bounds in turn.
Using the hyperbolic sub-interval~(\ref{hyper-data1}),
one can calculate from~(\ref{redshift-distR}) the largest 
values of $\chi_{obs}$ in this interval.
For $z_{max} = 6\,$ one finds $\chi^{max}_{obs} = 0.20125$,
while for  $z_{max} = 1100$ (CMBR) one finds  
$\chi^{max}_{obs} = 0.33745$.
Thus, using quasars up to $z_{max}=6\,$, FLRW hyperbolic 
universes with the density parameters in~(\ref{hyper-data1}) 
have undetectable topologies if their corresponding 
injectivity radii are such that $r_{inj} \geq  0.20125\,$. 
Similarly, for the same hyperbolic sub-interval, using CMBR, 
the  topology of hyperbolic universes with 
$r_{inj} \geq 0.33745$ is undetectable.
Further, for $z_{max} = \infty,$ the largest value of
$\chi_{obs}$ in the sub-interval~(\ref{hyper-data1}) is 
$\chi_{hor} = 0.349247$, so the  topology of hyperbolic 
universes with $r_{inj} \geq 0.34924$ is 
definitely undetectable regardless of depth of the survey.
In Table~1 we have summarized the  
restrictions on detectability imposed by the hyperbolic 
sub-interval~(\ref{hyper-data1}) on the first seven 
manifolds of Hodgson-Weeks census, where
$\,U$ denotes that the topology is 
undetectable by any survey of depth up to the redshifts 
$z_{max}=6$ (quasars) or  $z_{max}=1100$ (CMBR) respectively. 
Thus using quasars, the topology of the five known smallest 
hyperbolic manifolds, as well as m009(4,1), are undetectable 
within the hyperbolic region of the parameter space given
by~(\ref{hyper-data1}), while only topologies  m007(3,1) and 
m009(4,1) remain undetectable even if CMBR observations 
are used. This shows clearly how detectability depends 
concretely on the survey used.

\begin{table}[!ht]
\begin{center}
\begin{tabular}{*{4}{|c}|} \hline
$M$ & $r_{inj}$ & {\sc quasars} & {\sc cmbr}  \\ \hline \hline 
 m003(-3,1) & 0.292 & $U$ & --- \\ \hline
 m003(-2,3) & 0.289 & $U$ & --- \\ \hline
 m007(3,1)  & 0.416 & $U$ & $U$ \\ \hline
 m003(-4,3) & 0.287 & $U$ & --- \\ \hline
 m004(6,1)  & 0.240 & $U$ & --- \\ \hline
m004(1,2)   & 0.183 & --- & --- \\ \hline
 m009(4,1)  & 0.397 & $U$ & $U$  \\ \hline
\end{tabular}
\caption{ \label{Tb:HW-Census1} 
\small 
Restrictions on detectability of cosmic topology
originated from the sub-interval~(\ref{hyper-data1}) 
for the first seven manifolds of Hodgson-Weeks census. 
Here $U$ stands for undetectable using 
catalogues of quasars (up to $z_{max}=6$) or  CMBR
($z_{max}=1100$).}
\end{center}
\end{table}

Considering now the second hyperbolic sub-interval%
~(\ref{hyper-data2}), one can again calculate 
{}from~(\ref{redshift-distR})
the largest values of $\chi_{obs}$ in this 
interval: for $z_{max} = 6\,$ one has 
$\chi^{max}_{obs} = 0.313394$,
while for  $z_{max} = 1100$ (CMBR) one finds  
$\chi^{max}_{obs} = 0.538276$.
Thus, using quasars up to $z_{max}=6\,$, FLRW hyperbolic 
universes with the density parameters in~(\ref{hyper-data2}) 
have undetectable topologies if $r_{inj} \geq  
0.313394\,$. 
Similarly, for the same hyperbolic sub-interval, using CMBR, 
the  topology of hyperbolic universes with 
$r_{inj} \geq 0.538276$ is undetectable.
Further, for $z_{max} = \infty,$ the largest value of
$\chi_{obs}$ in the sub-interval~(\ref{hyper-data1}) is 
$\chi_{hor} = 0.557832$, so the  topology of hyperbolic 
universes with $r_{inj}$ greater than this value is 
definitely undetectable regardless of the depth of the survey.
The restrictions on detectability imposed by the hyperbolic 
sub-interval~(\ref{hyper-data2}) on the first seven 
manifolds of Hodgson-Weeks census can again be reexamined.
Using this sub-interval we find a very different picture from that 
summarized in Table~1, namely that in this case, using quasars, 
only two topologies [$\,m007(3,1)$ and $m009(4,1)\,$] would be 
undetectable whereas using CMBR none of topologies of these seven 
manifolds (universes) would be undetectable.
Table~2 summarizes the restrictions on detectability imposed 
by the hyperbolic sub-interval~(\ref{hyper-data2}) on the first 
seven manifolds of Hodgson-Weeks census. 
%
\begin{table}[!ht]
\begin{center}
\begin{tabular}{*{4}{|c}|} \hline
$M$ & $r_{inj}$ & {\sc quasars} & {\sc cmbr}  \\ \hline \hline 
 m003(-3,1) & 0.292 & --- & --- \\ \hline
 m003(-2,3) & 0.289 & --- & --- \\ \hline
 m007(3,1)  & 0.416 & $U$ & --- \\ \hline
 m003(-4,3) & 0.287 & --- & --- \\ \hline
 m004(6,1)  & 0.240 & --- & --- \\ \hline
 m004(1,2)  & 0.183 & --- & --- \\ \hline
 m009(4,1)  & 0.397 & $U$ & --- \\ \hline
\end{tabular}
\caption{ \label{Tb:HW-Census2} 
\small 
Restrictions on detectability of cosmic topology
originated from the sub-interval~(\ref{hyper-data2}) 
for the first seven manifolds of Hodgson-Weeks census. 
Here $U$ stands for undetectable using 
catalogues of quasars (up to $z_{max}=6$) or CMBR 
($z_{max}=1100$).}
\end{center}
\end{table}

The results in Table~2 together with those in Table~1   
make transparent that a variation of $1\%$ in the total
density parameters $\Omega_0$ ($0.99 \to 0.98$), for
$\Omega_{\Lambda 0} \in [0.62, 0.79]\,$,  which would
have no significant consequences in the geometrical (dynamical)
features of the universes, would crucially change the
detectability of cosmic topology.

One can also reexamine, in the light of the new bounds~(\ref{hyper-data2}),
what is the region of the parameter space for which a given set of 
topologies are undetectable. 
To this end we note that for a given topology (fixed $r_{inj}$) 
and for a given survey up to $z_{max}$, one can solve the 
equation
\begin{equation} \label{XobsRinj}
\chi^{}_{obs} (\Omega, \Omega_{\Lambda})\, = \, r_{inj} \;,
\end{equation}
which amounts to finding pairs ($\Omega,\Omega_{\Lambda}$)
in the density parameter plane for which Eq.~(\ref{XobsRinj}) 
holds.

Consider now the set of the 19 smallest manifolds of the 
Hodgson-Weeks census in conjuction with the hyperbolic 
region~(\ref{hyper-data2}) and the eqs.~(\ref{redshift-distR})
and~(\ref{XobsRinj}). The manifold in this set with the 
lowest $r_{inj} (= 0.152)$ is $m003(-5,4)\,$.
\begin{figure}[!htb]
\centerline{\def\epsfsize#1#2{0.5#1}\epsffile{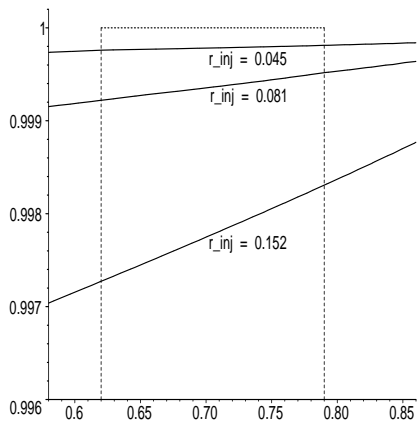}}
\caption{\label{fig=0.02} The solution curves of $\chi_{obs} = r_{inj}$, 
as plots of $\Omega_0$ (vertical axis) versus $\Omega_{\Lambda 0}$ 
(horizontal axis), for $r_{inj}=0.045$ (upper curve),$r_{inj}=0.081$ 
and $r_{inj} = 0.152$ (lower curve). A survey with  depth 
$z_{max}= 1100$ (CMBR)was used in both cases. 
The dashed rectangular box represents the relevant part, for our 
purposes, of the hyperbolic region~(\ref{hyper-data2}) of the 
parameter space given by recent observations.   
The undetectable regions of the parameter space 
($\Omega_0, \Omega_{\Lambda 0}$), corresponding to each 
value of $ r_{inj}$, lie above the related curve.}
\end{figure}
%
%
%
Figure~\ref{fig=0.02} gives the solution curve of 
equation~(\ref{XobsRinj}) in the 
$\Omega_0\,$--$\,\,\Omega_{\Lambda 0}$ plane for 
$r_{inj}=0.152$ and $r_{inj}= 0.081$, 
where a survey of depth $z_{max}=1\,100$ (CMBR) was used.%
\footnote{
This figure updates Figure~1 of in~\cite{grt2001b} 
two regards: first it employs a more recent hyperbolic sub-interval 
of the cosmological density parameters~\cite{Jaffe}; 
second it uses the most recent lower bound on the length of shortest 
closed geodesic in closed orientable hyperbolic 3-manifolds~\cite{HodKer}. 
Note that as opposed to the manifolds considered in~\cite{grt2001b} 
which could contain only Weeks' manifold, now there are at least 
19 manifolds.}
This figure also contains a dashed rectangular box, 
representing the relevant part (for our purposes here) 
of the recent hyperbolic region~({\ref{hyper-data2}).
For each value of $r_{inj}$ undetectability is ensured 
for the values of cosmological parameters (region in 
the $\Omega_{0}\,$--$\,\,\Omega_{\Lambda 0}$ plane) 
which lie above the corresponding solution curve 
of~(\ref{XobsRinj}).
Thus considering the solution curve of~(\ref{XobsRinj}) for 
$r_{inj} = 0.081$, one finds that all closed orientable 
hyperbolic manifolds (universes) with volumes less than 
$1.0711$, for example, would have undetectable topology, if 
the total density $\Omega_0$ turned out to be higher than 
$\sim 0.9994$. 
Similarly, considering the solution curve of~(\ref{XobsRinj}) 
for $r_{inj} = 0.152$, for example, one finds that the topology 
of none of the 19 smallest manifolds of the census would be 
detectable, if $\Omega_0$ turned out to be higher 
than $\sim 0.9974$. 

\section{Conclusions}

Motivated by the recent
observational results indicating that the universe is
nearly flat, we have employed the recent analyses of the
observational constraints on the cosmological density
parameters, together with recent mathematical results
concerning small hyperbolic $3$-manifolds, to examine the
possibility of detecting the topology of nearly flat
hyperbolic universes by using patterns repetition.
In this way we have updated and extended our recent results, 
which has resulted in new bounds on detectable topologies.

In addition we have also found that small changes in the 
cosmological density parameters of the order a few percent are 
sufficient to radically effect the detectability of the topology 
of small hyperbolic universes. This result, which is essentially 
the consequence of the rapid way the horizon radius $\chi_{hor}$
falls off to zero for nearly flat hyperbolic universes, is
of great potential importance, as it demonstrates concretely how
small changes in the observational bounds on the cosmological 
density parameters could have important consequences for the 
question of detectability of the cosmic topology.

\section*{Acknowledgments}
We are grateful to Ian Agol and Andrew Przeworski for very 
helpful correspondence concerning their work, and for 
drawing our attention to the~\cite{HodKer}. 
We also thank CNPq, MCT/CBPF and CLAF for the grants under 
which this work was carried out.




\begin{thebibliography}{99}

%
\bibitem{BooMax} P. de Bernadis {\em et al.\/}, {\em Nature}
{\bf 404}, 955 (2000); \\ 
%
S. Hanany {\em et al.\/}, \emph{Astrophys.\ J. 
Lett.\/} {\bf 545}, 5 (2000);  \\   
%
A.E. Lange {\em et al.\/}, \emph{Phys.\ 
Rev.\ D\/} {\bf 63}, 042001 (2001); \\  
%
P. de Bernardis {\em et al.\/}, \emph{First Results from 
BOOMERANG Experiment\/}, astro-ph/0011469 (2000). In 
Proc.\ of the CAPP2000  conference, Verbier, 17-28 
July 2000; \\
%
J.R. Bond {\em et al.\/}, \emph{The Cosmic Background 
Radiation circa $\nu$2K\/}, astro-ph/0011381 (2000). 
In Proc.\ of Neutrino 2000 (Elsevier), CITA-2000-63, 
Eds. J. Law \& J. Simpson; \\ 
%
J.R. Bond {\em et al.\/},  \emph{The Quintessential CMB, 
Past \& Future\/}. In Proc.\ of CAPP-2000 (AIP), 
CITA-2000-64; \\
%
J.R. Bond {\em et al.\/}, \emph{CMB Analysis of Boomerang 
\& Maxima \& the Cosmic Parameters\/}, 
astro-ph/0011378 (2000). In Proc. IAAU Symposium 201 (PASP), 
CITA-2000-65; \\
%
A. Balbi {\em et al.\/}, \emph{Astrophys. J.\/} {\bf 545}, 
L1--L4 (2000). 
%
%
\bibitem{Jaffe} A.H. Jaffe {\em et al.\/}, \emph{Phys.\ 
Rev.\ Lett.\/} {\bf 86}, 3475 (2001).
%
%
\bibitem{CosmicTop} 
G.F.R. Ellis, \emph{Gen.\ Rel.\ Grav.\/} {\bf 2}, 7 (1971); \\
%
D.D. Sokolov \& V.F. Shvartsman, \emph{Sov.\ Phys.\ JETP\/}
{\bf 39}, 196 (1974); \\
%
G.F.R. Ellis \& G. Schreiber, \emph{Phys.\ Lett.\ A\/} {\bf 115},
97 (1986); \\
%
R. Lehoucq, M. Lachi\`eze-Rey \& J.-P.
Luminet, \emph{Astron.\ Astrophys.\/} {\bf 313}, 339
(1996);\\
%
B.F. Roukema, \emph{Mon.\ Not.\ R.
Astron.\ Soc.\/} {\bf 283}, 1147 (1996); \\
%
G.F.R. Ellis \& R. Tavakol, \emph{Class.\ Quantum Grav.\/}
{\bf 11}, 675 (1994); \\
%
J.J. Levin, J.D. Barrow \& J. Silk,
\emph{Phys.\ Rev.\ Lett.\/} {\bf 79}, 974 (1997); \\
%
N.J. Cornish, D.N. Spergel \& G.D. Starkman, 
\emph{Class.\ Quantum Grav.\/} {\bf 15}, 2657 (1998); \\
%
N.J. Cornish, D.N. Spergel \& G.D. Starkman, 
\emph{Proc.\ Nat.\ Acad.\ Sci.\/} {\bf 95}, 82 (1998); \\
%
N.J. Cornish, D. Spergel \& G. Starkman, 
\emph{Phys.\ Rev.\ D\/} {\bf 57}, 5982 (1998); \\
%
J.J. Levin, E. Scannapieco \& J. Silk,
\emph{Phys.\ Rev.\ D\/} {\bf 58}, 103516 (1998); \\
%
J.J. Levin, E. Scannapieco \& J. Silk,
\emph{Class.\ Quantum Grav.\/} {\bf 15}, 2689 (1998); \\ 
%
%
J.J. Levin, E Scannapieco, E. Gasperis, J. Silk,
\& J.D. Barrow, \emph{Phys.\ Rev.\ D\/} {\bf 58}, 123006 
(1999); \\
%
R. Lehoucq, J.-P. Luminet \& J.-P. Uzan,
\emph{Astron.\ Astrophys.\/} {\bf 344}, 735 (1999); \\
%
J.-P. Uzan, R. Lehoucq \& J.-P. Luminet,
\emph{Astron.\ Astrophys.\/} {\bf 351}, 776 (1999); \\
%
H.V. Fagundes \& E. Gausmann,
\emph{Phys.\ Lett.\ A\/} {\bf 261}, 235 (1999); \\
%
R. Aurich, \emph{Astrophys. J.\/} {\bf 524}, 497 (1999); \\
%
J.R. Bond, D. Pogosyan \& T. Souradeep, 
\emph{Phys.\ Rev.\ D\/} {\bf 62}, 043005 (2000); \\
%
J.R. Bond, D. Pogosyan \& T. Souradeep, 
\emph{Phys.\ Rev.\ D\/} {\bf 62}, 043006 (2000); \\
%
G.I. Gomero, M.J. Rebou\c{c}as \& A.F.F. Teixeira,
\emph{Phys.\ Lett.\ A\/} {\bf 275}, 355 (2000); \\
%
G.I. Gomero, M.J. Rebou\c{c}as \& A.F.F. Teixeira, 
\emph{Int.\ J. Mod.\ Phys.\ D\/} {\bf 9}, 687 (2000); \\
%
G.I. Gomero, M.J. Rebou\c{c}as \& A.F.F. Teixeira,
\emph{Class.\ Quantum Grav.\/} {\bf 18}, 1885 (2001); \\ 
%
M.J. Rebou\c{c}as,
\emph{Int.\ J. Mod.\ Phys.\ D\/} {\bf 9}, 561 (2000); \\
%
G.I. Gomero, A.F.F. Teixeira, M.J. Rebou\c{c}as 
\&  A. Bernui, \emph{Int.\ J.\ Mod.\ Phys.\ D} {\bf 11},
869 (2002). \\  
%
G.I. Gomero \& M.J. Rebou\c{c}as, \emph{Detectability of 
Cosmic Topology in  Flat Universes}, gr-qc/0203094 (2002). \\
%
R. Aurich \& F. Steiner, \emph{Mont.\ Not. Roy.\ Astron.\
Soc.\/} {\bf 323}, 1016 (2001). 

%
\bibitem{Revs} 
J.J. Levin, \emph{Phys.\ Rep.\/} {\bf 365}, 251 (2002).\\
%
R. Lehoucq, J.-P. Uzan \& J.-P. Luminet,
astro-ph/0005515 (2000); \\ 
%
V. Blanl{\oe}il \& B.F. Roukema, Eds., astro-ph/0010170 (2000); \\
%
G.D. Starkman, \emph{Class.\ Quantum Grav.\/}
{\bf 15}, 2529 (1998); \\ 
%
M. Lachi\`eze-Rey \& J.-P. Luminet,
\emph{Phys.\ Rep.\/} {\bf 254}, 135 (1995).  

\bibitem{ZelNov83} Ya. B. Zeldovich \& I.D. Novikov, {\em The 
Structure and Evolution of the Universe\/}, p.~633-640, 
The University of Chicago Press (1983). See on p.~637 
refs.\ of the earlier works by
S\"{u}veges (1966), Sokolov (1970), Pa\'al (1971), 
Sokolov and Shvartsman (1974), and Starobinsky (1975).


%
%
\bibitem{Supernovae} B.P. Schmidt {\em et al.\/}, 
\emph{Astrophys.\ J.\/} {\bf 507} 46 (1998); \ 
%
A.G. Riess {\em et al.\/}, \emph{Astron.\ J.\/} {\bf 116}, 
1009 (1998).
\bibitem{RoukMamBaj02} B.F. Roukema, G.A. Mamon \& S. Bajtlik,
\emph{Astron.\ Astrophys.\/} {\bf 382}, 397 (2002).


%
%
\bibitem{Acc} S. Perlmutter {\em et al.\/}, \emph{Astrophys. J.\/} 
{\bf 517}, 565 (1999); \  
%
S. Perlmutter, M.S. Turner \&  M. Write, \emph{Phys.\ Rev. Lett.\/} 
{\bf 83}, 670 (1999).
%
\bibitem{grt2001a} G.I. Gomero, M.J. Rebou\c{c}as \& R. Tavakol,
\emph{Class.\ Quantum Grav.\/} {\bf 18}, 4461 (2001).

\bibitem{grt2001b} G.I. Gomero, M.J. Rebou\c{c}as \& R. Tavakol,
\emph{Class.\ Quantum Grav.\/} {\bf 18}, L145 (2001).

\bibitem{EvLeLuUzWe} E. Gausmann, R. Lehoucq, J-P Luminet,
J-P Uzan, J. Weeks, {\em Class. Quantum Grav.} {\bf 18}, 
5155 (2001).

\bibitem{Mostow} G.D. Mostow,  {\em Ann.\ Math.\ Studies\/} 
{\bf 78} (1973), Princeton University Press, Princeton, 
New Jersey.  

\bibitem{Thurston82} W.P. Thurston, {\em Bull.\ Am.\ Math.\ 
Soc.\/} {\bf 6}, 357 (1982).

\bibitem{Agol} I. Agol \emph{Volume Change under Drilling\/},
preprint available at 
{\tt http://xxx.lanl.gov/abs/math.GT/0101138\/} (2001).

\begin{sloppypar}
\bibitem{Przeworski} A. Przeworski,  {\em J. Differential Geom.\/}
{\bf 58}, 2 (2001). Also available at 
{\tt http://www.ma.utexas.edu/users/prez/\/}.
\end{sloppypar}

\bibitem{SnapPea} J.R. Weeks, SnapPea: A computer program
for creating and studying hyperbolic 3-manifolds, available at 
{\tt http://thames.northnet.org/weeks/}

\bibitem{AdamsSnap} C. Adams, {\em Not.\ Am.\ Math.\ Soc.\/} 
{\bf 37}, 273 (1990).

\bibitem{HodgsonWeeks} C.D. Hodgson \& J.R. Weeks,
\emph{Experimental Mathematics\/} {\bf 3}, 261 (1994).

\bibitem{HodKer} C.D. Hodgson \& Steven P. Kerckhoff,
\emph{Universal Bounds for Hyperbolic Dehn Surgery\/},
preprint available at 
{\tt http://xxx.lanl.gov/abs/math.GT/0204345\/}.(2002).

\bibitem{Bond-et-al-00a} J.R. Bond {\em et al.\/},  
\emph{The Quintessential CMB, Past \& Future\/}. In
Proc.\ of CAPP-2000 (AIP), CITA-2000-64.


\end{thebibliography}
\end{document}